\documentclass[apjl]{emulateapj}
\usepackage{graphicx}
\usepackage{cancel}
\usepackage{tikz}
\usepackage{amsmath,amssymb}
\usepackage{subfigure}
\usepackage{afterpage}

\DeclareMathOperator{\cov}{cov}

\newcommand{\e}[1]{\times10^{#1}}
\newcommand{\nc}{\newcommand}

\newcommand{\Epsilon}{\mathcal{E}}
\nc{\hf}{\frac{1}{2}}
\nc{\hfrt}{\frac{1}{\sqrt{2}}}
\nc{\fb}[2]{\left(\frac{#1}{#2}\right)}
\nc{\sqb}[2]{\sqrt{\frac{#1}{#2}}}
\nc{\expb}[1]{\exp\left[#1\right]}

\nc{\twovec}[2]{\left(\begin{array}{c}#1\\#2\end{array}\right)}
\nc{\abs}[1]{\left|#1\right|}
\nc{\vc}{\vec{\sigma}}
\nc{\degree}{^{\circ}}

\slugcomment{Accepted by the Astrophysical Journal}

\begin{document}
\title{Comparison of Diversity of Type IIb Supernovae with Asymmetry in Cassiopeia A Using Light Echoes}
\author{Kieran Finn$^1$, Federica B. Bianco$^{1,2}$,  Maryam Modjaz$^1$, Yu-Qian Liu$^1$, Armin Rest$^3$} 
\affil{$^1$Center for Cosmology and Particle Physics, New York University, New York, NY 10003, USA}
\affil{$^2$
Center for Urban Science and Progress, New York University,
1 MetroTech Center, Brooklyn, NY 11201, USA}
\affil{$^3$Space Telescope Science Institute, 3700 San Martin Drive, Baltimore, MD 21218, USA}

\begin{abstract}
  We compare the diversity of spectral line velocities in a large sample of type IIb supernovae (SNe~IIb) with the expected asphericity in the explosion, as measured from the light echoes of Cassiopeia A (Cas~A), which was a historical galactic SN~IIb. We revisit the results of \citet{Rest:2010xc}, who used light echoes 
  to observe Cas~A from multiple lines of sight and hence determine its asphericity, as seen in the velocity of three spectral lines (He I $\lambda$5876, H$\alpha$ and the Ca II NIR triplet). We confirm and improve on this measurement by reproducing the effect of the light echoes in the spectra of several extragalactic SNe~IIb found in the literature as well as mean SN~IIb spectra recently created by \citet{Yuqian}, and comparing these to the observed light echo spectra of Cas~A, including their associated uncertainties. In order to quantify the accuracy of this comparison we smooth the light echo spectra of Cas~A using Gaussian processes and use a Monte Carlo method to measure the absorption velocities
 of these three features in the spectra. We then test the hypothesis that the diversity of ejecta velocities seen in SNe~IIb can be explained by asphericity. We do this by comparing the range of velocities seen in the different light echoes, and hence different lines of sight, of Cas~A to that seen in the population of SNe~IIb. 
We conclude that these two ranges are of the same order and thus asphericity could be enough to explain the diversity in the expansion velocity alone.
\end{abstract}

\section{Introduction}
\label{sec:intro}

A large number of supernovae (SNe) have been detected and analyzed, leading to a sophisticated classification scheme \citep{filippenko}, which is believed to describe the diversity of the progenitors and explosion conditions. Within this scheme, stripped-envelope SNe \citep{Clocchiatti97,Filippenko97,Matheson01,Modjaz:2014doa} constitute a subclass containing about 25\% \citep{Li11}
of all SNe. They occur when a massive star loses some or all of its hydrogen and helium layers before suffering a core-collapse. There are several mechanisms by which the hydrogen and helium layers can be removed, for example
they can be ejected by eruptions, gradually stripped by strong winds \citep[e.g.][]{Woosley93} or binary interactions \citep{Nomoto95, Podsiadlowski04}, or
burned in the presence of enhanced mixing \citep{Yoon:2005vx, Frey13}. Stripped-envelope SNe are identified
observationally by the weakness or absence of hydrogen spectral features (SN~IIb and SN~Ib respectively), especially at late phases, and by the absence of hydrogen and helium features (SN~Ic and SN~Ic-bl). The progressive lack of hydrogen and helium in the spectra is
the signature of the stripping.

The most extensive spectral data set of stripped-envelope SNe was recently published by \citet{Modjaz:2014doa}. \citet{Modjaz:2015cca} and \citet[henceforward L16]{Yuqian} showed that there
is a great deal of diversity, even within a specific subtype, including within type IIb Supernovae (SNe~IIb), which are the focus of our study. SNe~IIb have helium and a small amount of hydrogen in their pre-explosion progenitors. 
In particular, there is a large range of absorption velocities measured for several spectral features. L16 quantified this diversity by constructing mean SN~IIb spectra and analyzing the relative standard deviation from these data. This is the first instance of stripped-envelope SN templates in the literature, where the spectral mean behavior and diversity of SNe~IIb are systematically quantified.

The question as to the cause of this diversity still remains, but one possibility is that the phenomena from which SNe~IIb arise are intrinsically identical but asymmetric and thus the diversity comes from observing along different lines of sight. Asphericity has been observed in many SNe using e.g. polarization \citep{Maund:2007ag} and double peaked emission lines \citep[e.g.][]{Modjaz:2008si, Maeda:2008mw}, which makes this a plausible scenario. 
Another technique for determining asphericity uses light echoes (LEs,~\citealt{Couderc:1939ab,Crotts:1988ab}) to obtain ``3D spectroscopy''; this is particularly useful for our purpose since it allows us to observe the spectrum of a source from multiple lines of sight \citep{Boumis98,Rest:2010xc,Rest:2010yi,Sinnott13}. For long duration transients (such as the great eruption of $\eta$-Carine, but also possibly tidal disruption events, super luminous SN, and even for regular SNe in the presence of extremely thin reflecting dust filaments, see Section~\ref{sec:dust}), spectroscopic time series can be studied through LEs, assessing the spectroscopic evolution from different lines of sight~\citep{Prieto:2014, Rest:2014}.

A LE occurs when dust clouds in the vicinity of an object reflect some of the light towards earth, acting like a mirror. LEs are useful events for many reasons. 
Because the light must travel a greater distance to earth than the direct path, the time of flight delay allows us to observe historical SNe with modern equipment \citep{Rest:2005un,Rest:2008kw} and thus classify supernova remnants by their historical SN spectra.  Furthermore, if there are multiple LEs, we can observe the same object from multiple lines of sight and hence establish the degree of asymmetry within it \citep[henceforward R11]{Rest:2010xc}. We can then compare this asymmetry to the supernova remnant and to directly observed explosion properties.

To date there is only one stripped-envelope supernova for which spectra from multiple LEs have been obtained.\footnote{Multiple light echoes have been observed for SN~1993J \citep{Sugerman:2002dy,Liu:2002fp}, however no spectra were taken of these so we cannot apply our analysis.} Cassiopeia A (Cas~A) is a Galactic SN
whose light reached earth in 1681 $\pm$ 19 years \citep{Fesen:2006zma}. LEs of the explosions were discovered and studied \citep{Rest:2007ab, Rest:2008kw, Krause:2008dm} leading to its spectroscopic classification as a SN type IIb.
R11 presented and analyzed spectra of three Cas~A LEs. By comparing the absorption velocity of three atomic transition lines in the Cas~A spectra with those in the spectra of SN~1993J and SN~2003bg, two well observed SNe~IIb, they found that the lines in the Cas~A spectra exhibit different absorption velocities when observed from different lines of sight, which is a signature of asymmetry in the explosion. R11 noticed but did not quantify the velocity differences in the light echoes.
In this work, we wish to quantify this observation of diversity and its statistical significance.
Cas~A provides a unique opportunity to compare the explosion asymmetry of a single object seen from multiple lines of sight to the diversity of many objects, each seen from a single line of sight, and hence determine if the former could be the cause of the latter. 

R11's study of its LEs provides the first and, to date, only \emph{direct} evidence for asymmetry in Cas~A, however asymmetry in the Cas~A remnant has been detected in several other studies using a variety of methods.
The direct detection of asymmetry in the explosion of Cas~A through its LEs, whose statistical significance we determine in the pages that follow, fits a convincing scenario of the asymmetry of the remnant, which strengthens and supports the result.
\citet{Delaney:2010} and \citet{Milisavljevic:2012qr} 
created a 3D picture of the remnant of Cas~A (in the IR and X-ray, and optical respectively), using Doppler tomography and proper motion to track the velocities in various parts of the ejecta. By making the assumption that all objects were moving away freely from a single point they were then able to reconstruct a 3D map  of the remnant.\footnote{An interactive version of this map, created by Dan Milisavljevic, is available online at www.cfa.harvard.edu/$\sim$dmilisav/casa-webapp.} The 3D structure shows that this remnant is far from spherically symmetric.
Compact ejecta features, dubbed ``knots'', discovered by \citet{Fesen:2001ab} and confirmed by \citet{Hwang:2004za} also reveal asymmetry. Although these knots are mostly isotropic, there is a small group along the southwestern limb moving significantly quicker than the rest. Further, by studying the ratio of $^{44}$Ti/$^{56}$Ni, \citet{Nagataki:1998kq} showed that it was higher than theoretically allowed for a symmetric explosion, and
\citet{Grefenstette:2014ab} showed that the distribution of $^{44}$Ti as mapped by NUSTAR \citep{Harrison:2013md} is not uniform.

In this paper we provide a statistical assessment of the direct asymmetry of Cas~A by comparing the LE spectra with recent spectral templates, and a comparison of the asymmetry of Cas~A to the spectral line velocity diversity in the population of SNe~IIb.

\section{Data}
\label{sec:data}
For our analysis we used the Cas~A light echo data collected and processed by R11, spectra and light curves of individual SNe~IIb from the literature, mean SN~IIb spectra from L16 and SN~IIb template light curves obtained from the data in \citet{Bianco:2014mna}. In the following subsections we describe these datasets.

\subsection{Light Echo Data}
Three LEs of Cas~A were discovered, and their spectra  collected and processed by \citet{Rest:2008kw} and R11. While we refer to those publications for a detailed description of the data, here we discuss relevant characteristics of the LE spectra.

The data were taken with the Low Resolution Imaging Spectrometer \citep{Oke:1995jc} on the 10m Keck I telescope and with the Blue Channel spectrograph \citep{Schmidt:1989ab} on the 6.5m MMT. The LEs were originally labeled LE2116, LE2521 and LE3923 by R11, however hereafter we refer to them by the direction of the line of sight of Cas~A that each one probes: SW, NW and NE, respectively. The light echoes are separated by: SW-NW~$\sim70\degree$, SW-NE~$\sim85.5\degree$ and NE-NW~$\sim109\degree$, thus truly probing different directions of the explosion. The native resolution of these spectra is around 7\AA\ and the dispersion is 1\AA\ for the NE and SW LEs and 2\AA\ for the NW LE.

For the NW LE, two spectra were taken by R11 at epochs separated by one day, and are analyzed separately in their work. In order to improve the signal-to-noise ratio (SNR) of our measurements, we combine these two spectra into a single spectrum in the wavelength range covered by both, namely $3192 \AA\leq\lambda\leq8400 \AA$.
We smooth (Section~\ref{sec:velocity}) and renormalize the spectra and use their average, and combine the errors in quadrature.

\begin{figure}\center{
		\subfigure{\includegraphics[width=\columnwidth]{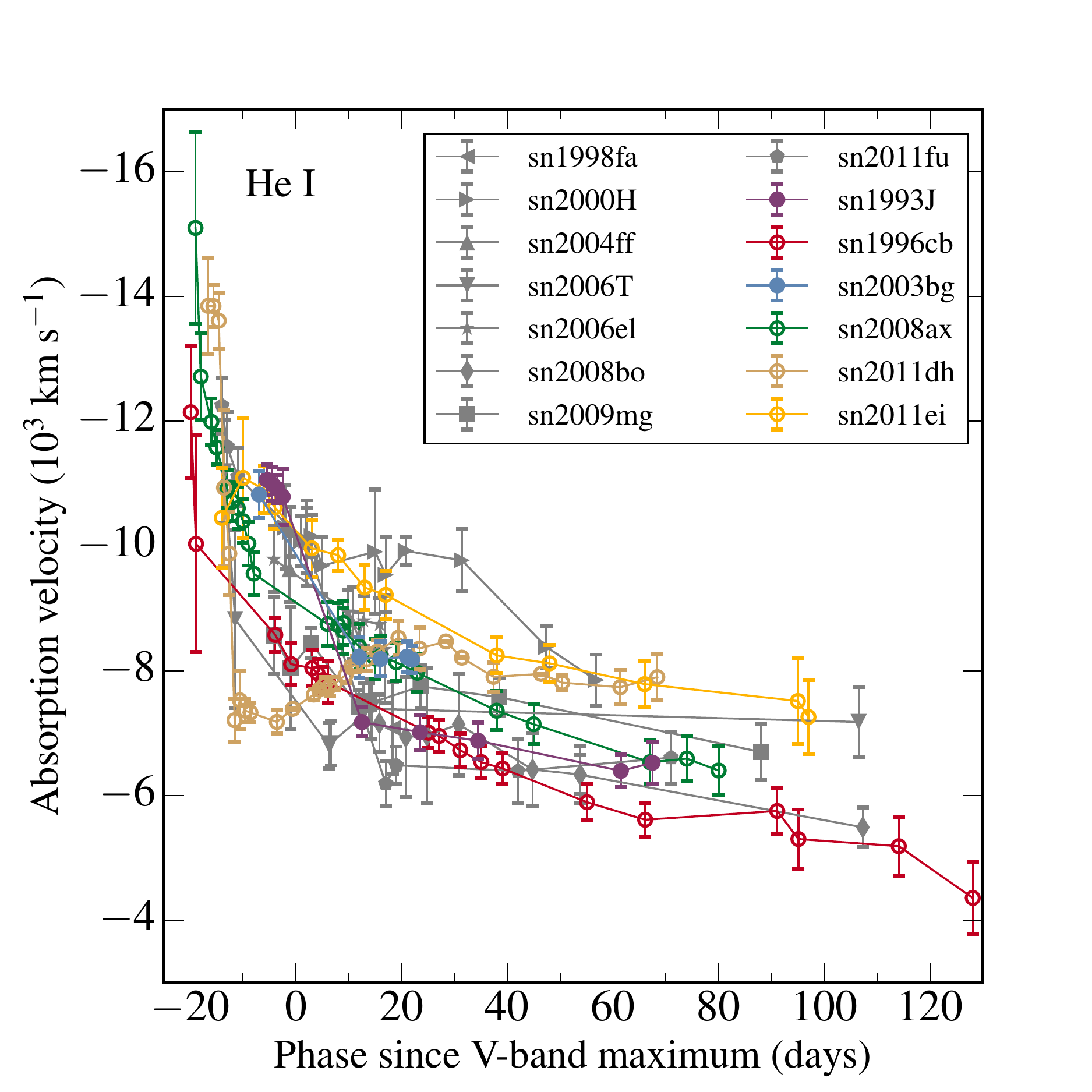}}\\
        \subfigure{\includegraphics[width=\columnwidth]{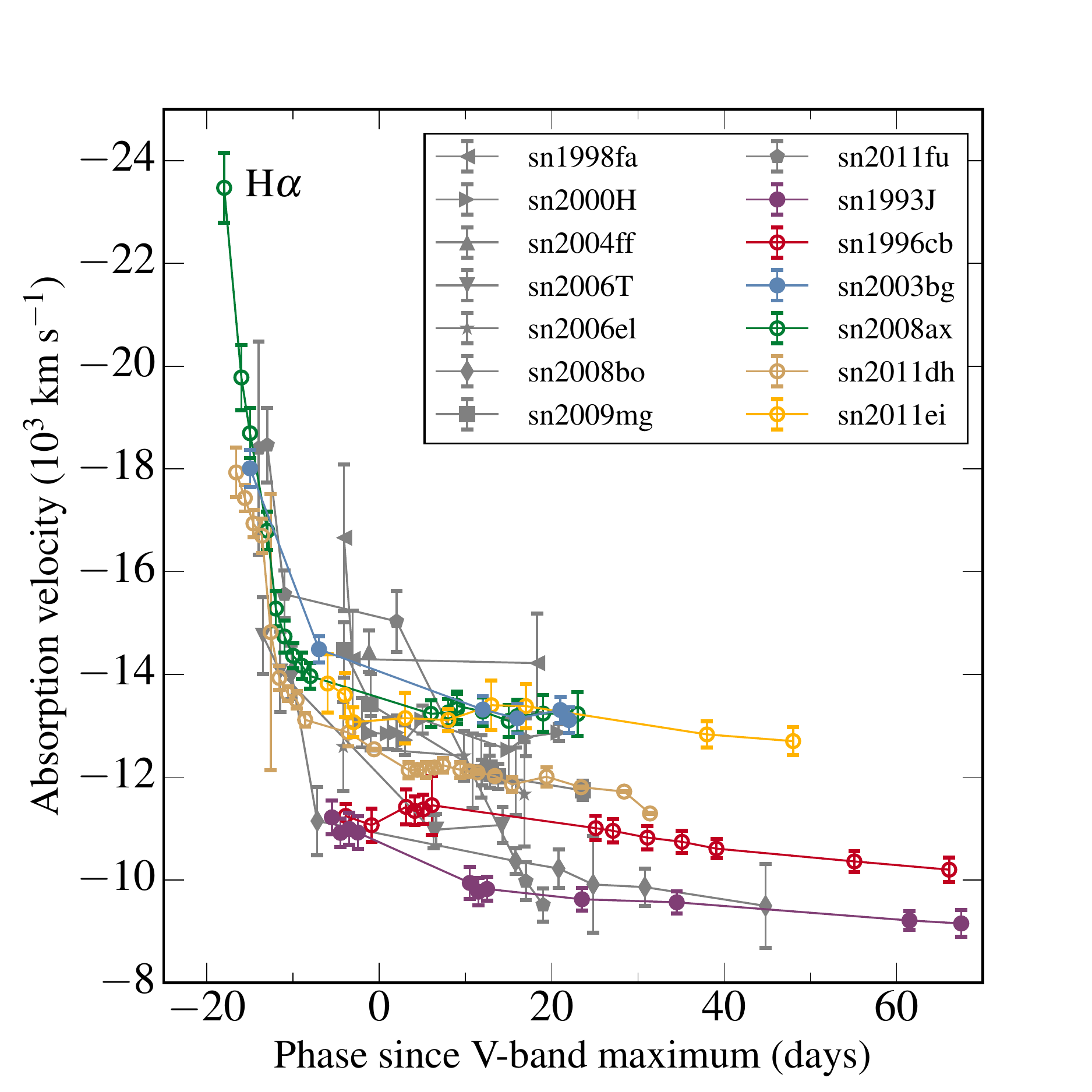}}}
        \caption{Evolution of velocities of He I $\lambda$5876 (upper panel) and H$\alpha$ (lower panel) for all SNe~IIb studied by L16 and from which the SN~IIb spectral template was generated. SN~1993J and SN~2003bg (solid symbols) were used in R11 as reference to investigate Cas~A's asymmetry. The SNe we use individually in our sample are SN~1993J, SN~1996cb, SN~2003bg, SN~2008ax, SN~2011dh, SN~2011ei, and are plotted in colors here. These SNe are representative of the range of velocities and velocity evolutions observed in the SNe~IIb population. Our sample is limited by the need for spectra at a range of phases with high SNR and the need for complete light curves. This is why e.g., SN~2000H and SN~1998fa are not included.\label{fig:yuqian_velocity}}
\end{figure} 

\subsection{Mean SN~IIb Spectra and Light Curves}

R11 measured and compared the velocity of the He I $\lambda$5876, H$\alpha$ and Ca II NIR triplet absorption features in the three LE spectra and in the spectra of SN~1993J and SN~2003bg. R11 chose these lines as a metric to measure asymmetry because they are strong, isolated lines in the spectra of SNe~IIb, and we expect their absorption velocity (measured as the location of the feature minimum) to approximately track the velocity of the outer envelopes of the star. By looking at this velocity in different directions we can build a picture of the asymmetry of the SN.

However, since the dust sheets are not perfect reflectors, one cannot compare the light-echo spectra directly to each other or to spectra from SNe in the literature, nor can one invert the effect of the dust to recover the emitted spectra. Instead one must first consider what effects the dust had on the Cas~A spectrum and then reproduce those effects in a chosen reference spectrum so that they are comparable. These modeling techniques are described in R11 and by several earlier works \citep{Couderc:1939ab,Chevalier:1986ab,Emmering:1989ab,Sugerman:2003ky,Patat:2005ab}. The modeling of the effect of reflection on spectra is described in \citet{Rest:2010yi} and we will use the 
same LE models, which are also used in \citet{Rest:2010xc}. An in-depth description of the method can be found in these works to which we refer the reader for details, but provide a brief summary in Section~\ref{sec:dust}.

\begin{figure*}
	\centerline{
		\subfigure{\includegraphics[width=\columnwidth]{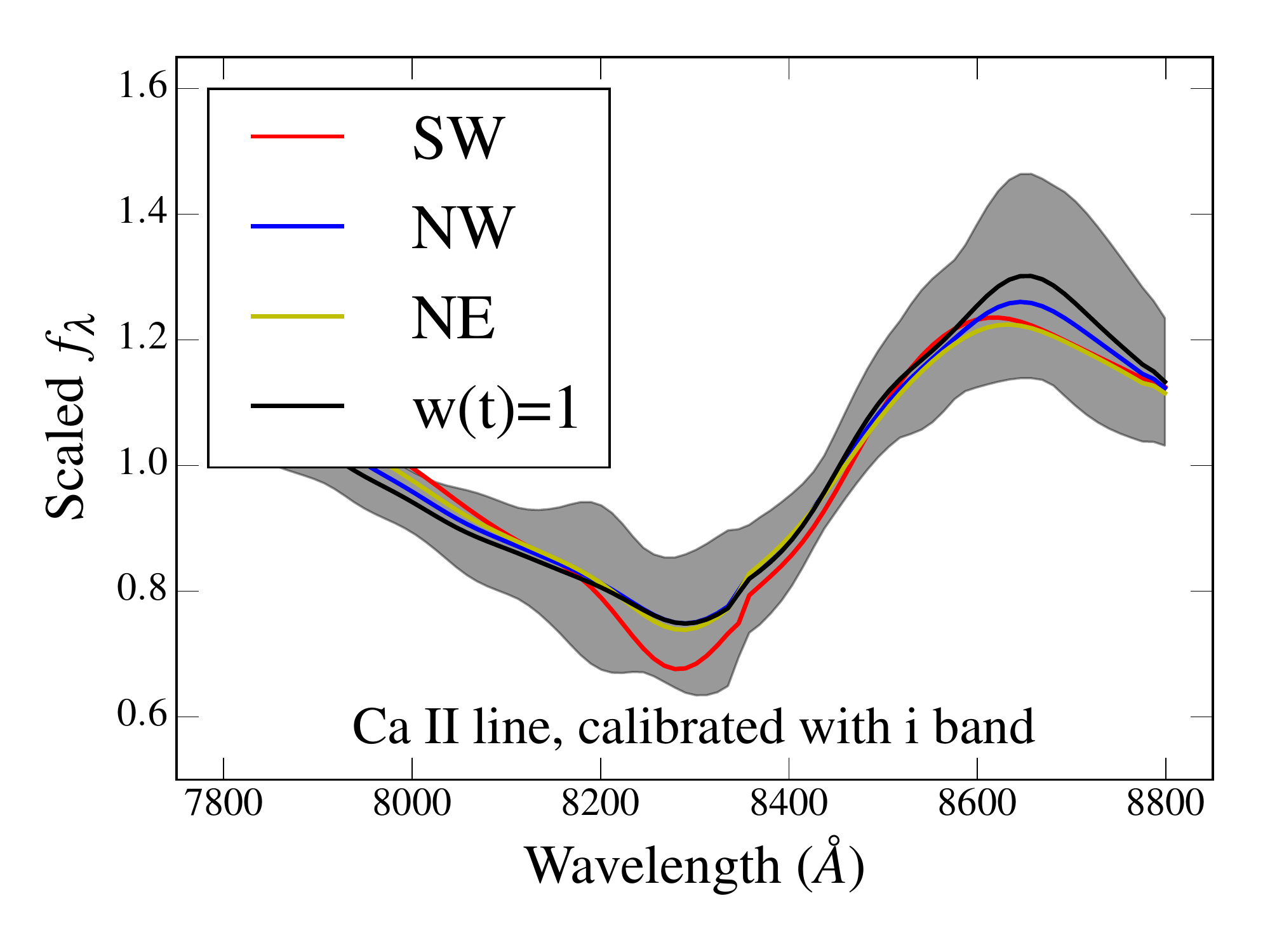}}
        \subfigure{\includegraphics[width=\columnwidth]{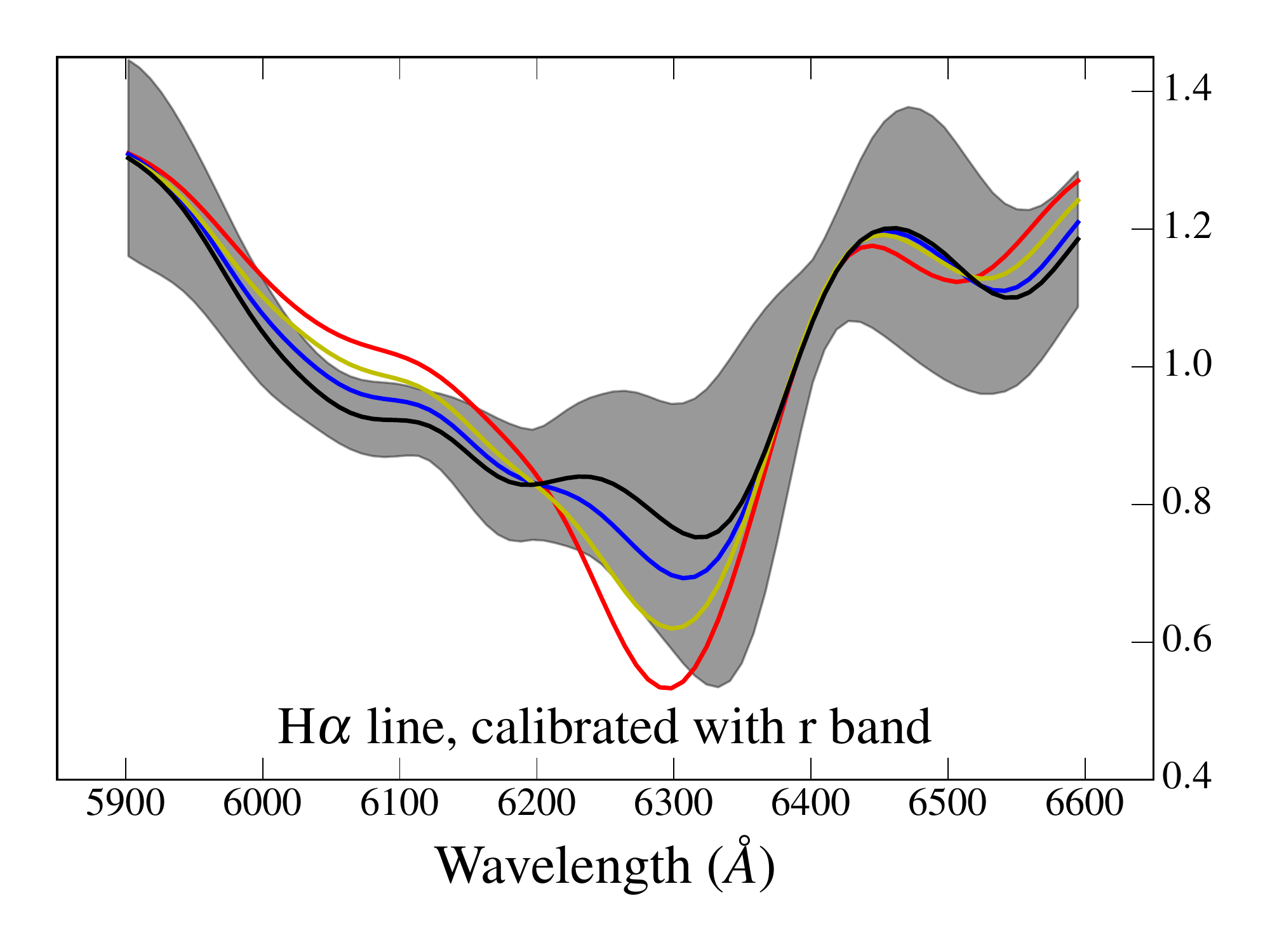}}}
        \centerline{
        \subfigure{\includegraphics[width=\columnwidth]{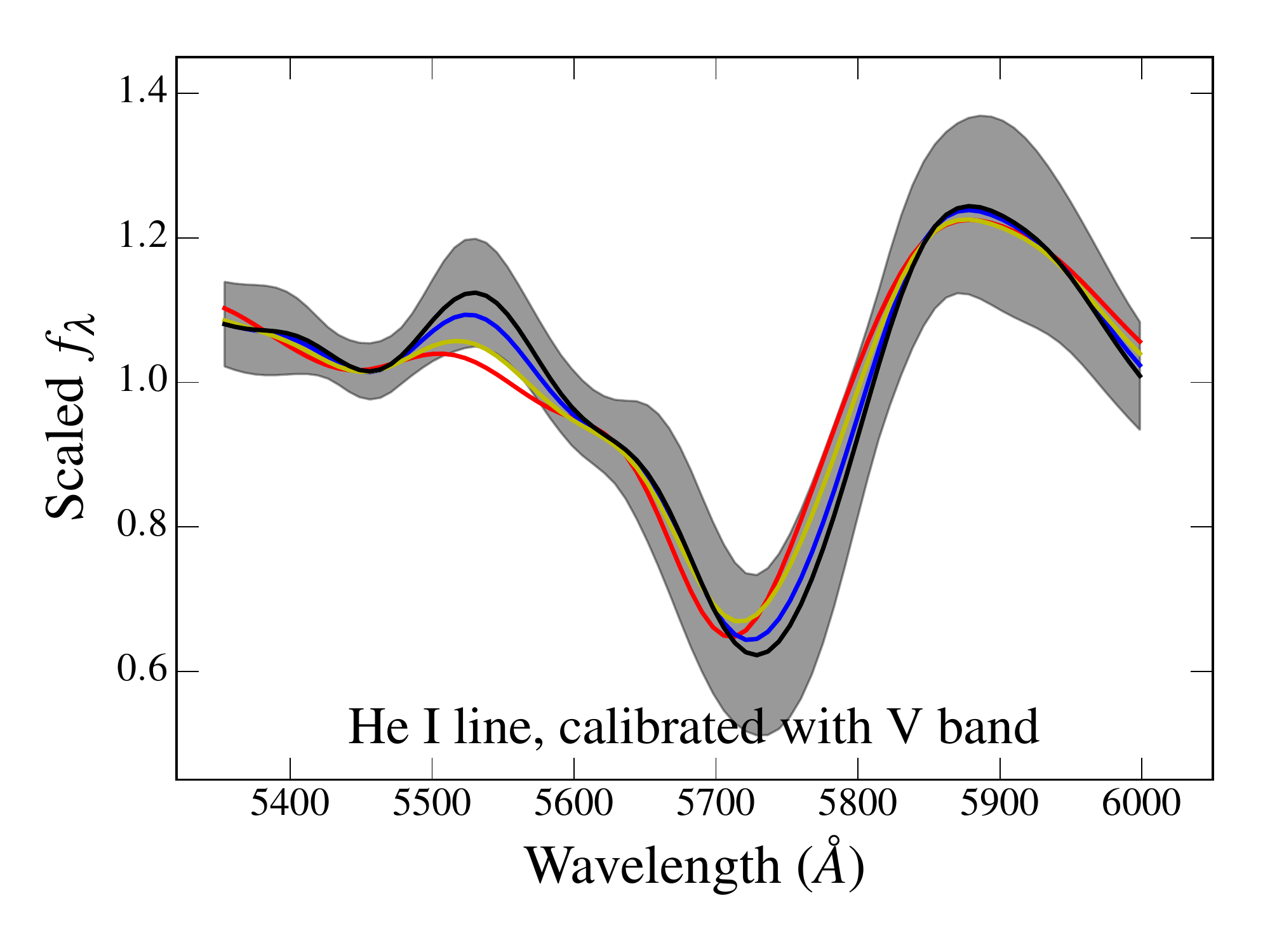}}
        \subfigure{\includegraphics[width=\columnwidth]{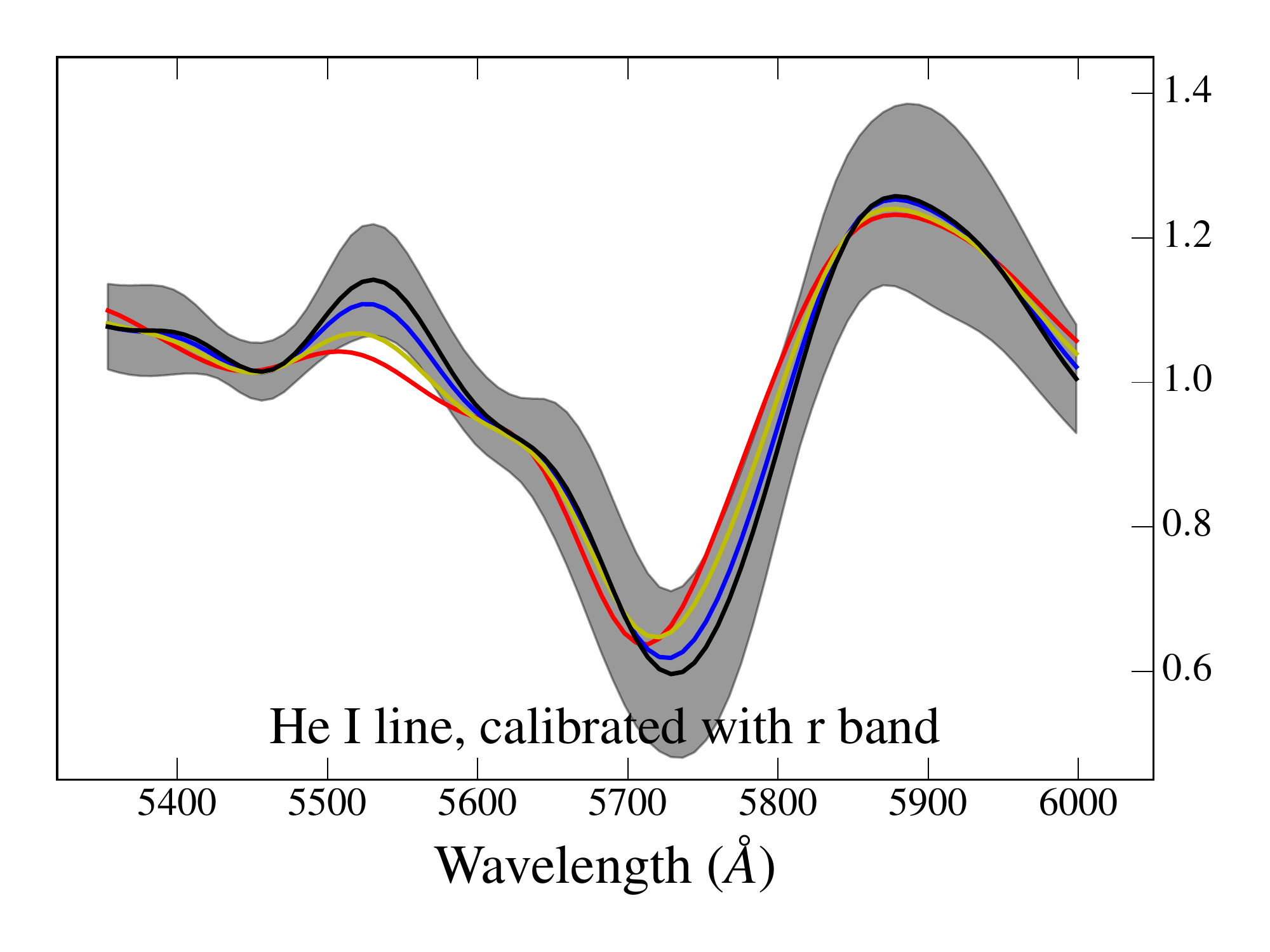}}}
        \caption{SN~IIb mean spectra from L16, and standard deviation in the sample (gray contours), in the wavelength region relevant for the Ca~II NIR triplet (top left), H$\alpha$ (top right), and He~I $\lambda$5876 spectral features (bottom left and right). The mean spectra are convolved with the window functions of the three Cas~A LEs, which includes the effects of dust, and integration over the photometric evolution. The $w(t)=1$ window function is simply the integration of the spectra at all epochs weighted by the photometry. The photometric band used for each feature is the band that best covers the wavelength range, and it is noted on the bottom of each plot. For the case of the He I $\lambda$5876 line, both V-band and r-band spectra were produced since this feature lies on the overlap of these two bands. The negligible difference between these two spectra justifies our use of a single band for each line and we use V-band normalization for the rest of the paper. For clarity, the standard deviation is shown only for the window function $w(t)=1$, however it is similar when processed through the LEs.}\label{fig:features}
\end{figure*}

In order to simulate the effect of dust scattering one must take both the spectral and photometric time evolution into account (Section~\ref{sec:dust}). Therefore, R11 chose the spectra of SN~1993J and SN~2003bg as the reference spectra since these SNe~IIb are both exceptionally well observed. However, analysis by L16 revealed that when it comes to our chosen metric, the absorption velocities of the three lines we are focusing on, these SNe are actually some of the most extreme, and have atypical evolution (e.g. $H\alpha$ for SN 1993J, see Figure~\ref{fig:yuqian_velocity}). We instead use a SN~IIb template that describes the average behavior of SNe~IIb as our reference spectra when evaluating the asymmetry of Cas~A (Figure~\ref{fig:features}). The SN~IIb spectral template was produced by L16 using 227 spectra of 14 SNe~IIb taken from the SuperNova IDentification code (SNID) spectra library \citep{Blondin:2007ua}, the sample presented in \citet{Modjaz:2014doa},
and relevant spectra from the literature available before September 2014. Template spectra, i.e. mean spectra, were created at phases ranging from $t_{Vmax}=-$20 to $+$45 in 5~day intervals and also at phases 55, 65, 70, 90, 95 and 105 by taking one spectrum from all SNe~IIb that had been observed within $\pm2$ days of the target phase and averaging. In addition, in order to retain a measure of the diversity of the SNe spectra that were used for constructing the mean spectra, we adopt the standard deviation spectral arrays provided by L16 and plot them as contours in our plots of the template spectra (Figure~\ref{fig:features}). Full details of how the template was generated can be found in L16.

We can use these template spectra to confirm the asymmetry of Cas~A, however we cannot use them to determine whether the asymmetry can explain the diversity of absorption velocities seen in SNe~IIb. To do this, we look at a selection of the individual SNe that make up the template. Our selection is chosen such that the complete range of absorption velocities is represented, as identified in L16, and in particular, those SNe that have the most extreme velocities are included in our sample. Figure~\ref{fig:yuqian_velocity} shows the velocity evolution of all SNe studied by L16 and was used to inform our choice. The SNe used, along with the references for their spectra and photometry are: SN~1993J \citep{Jeffery:1994ab,Barbon:1995ab,Richmond96,Matheson:2000ab,
Fransson:2005ab,Modjaz:2014doa}, SN~1996cb \citep{Qiu99,Matheson:2001ab,Modjaz:2014doa}, SN~2011dh \citep{Marion:2014ab}, SN~2003bg \citep{Hamuy:2009ab}, SN~2008ax \citep{Pastorello:2008uz,Milisavljevic:2010ab,Modjaz:2014doa,Bianco:2014mna},
SN~2011dh \citep{Marion:2014ab} and SN~2011ei \citep{Milisavljevic:2012qr}. Since the Ca triplet feature falls in near infrared (NIR) wavelengths, and NIR spectra of stripped SN are rare, the Ca line comparison cannot be conducted for all  SNe in our sample, and we restrict ourselves to only the He I $\lambda$5876 and H$\alpha$ lines for this part of the analysis.

\section{Methods}
\subsection{Effect of the Dust}
\label{sec:dust}

The dust sheets that cause the LEs do not act as perfect mirrors,
and thus the spectra we observe will not be the same as those emitted by the SN. One reason for this is that the dust sheets have a finite size and thickness, which means that light from multiple epochs in the event timeline will reach the observer at the same time by scattering off different parts of the dust. Therefore the spectrum we observe in a LE is really the sum of spectra from all epochs weighted appropriately with weights that depend on properties of the dust, of the observation, and on the photometric evolution of the transient, as opposed to that of a single epoch, which would be observed if viewing the SN directly.

Since the effect of this integration cannot be removed from the observed Cas~A spectra, we shall process a reference object in such a way as to simulate how it would have appeared had it exploded at the same position as Cas~A and been observed in the same LEs. It will then be possible to directly compare the spectra of the reference object and the Cas~A LE spectra. For this purpose a window function \citep{Tylenda:2003ty,Rest:2010yi}, $w(t)$, is defined for each LE such that the observed spectrum, $S(\lambda)$, is related to the emitted spectra, $s(\lambda,t)$, by

\begin{equation}
S(\lambda)=\int_{-\infty}^{\infty} s(t,\lambda)w(t)dt.
\end{equation}
It should be noted that the relative normalization of the different epochs that contribute to $s(t,\lambda)$ is important and thus knowledge of the photometry of the reference object is essential.

\begin{figure}[h]
\includegraphics[width=0.5\textwidth]{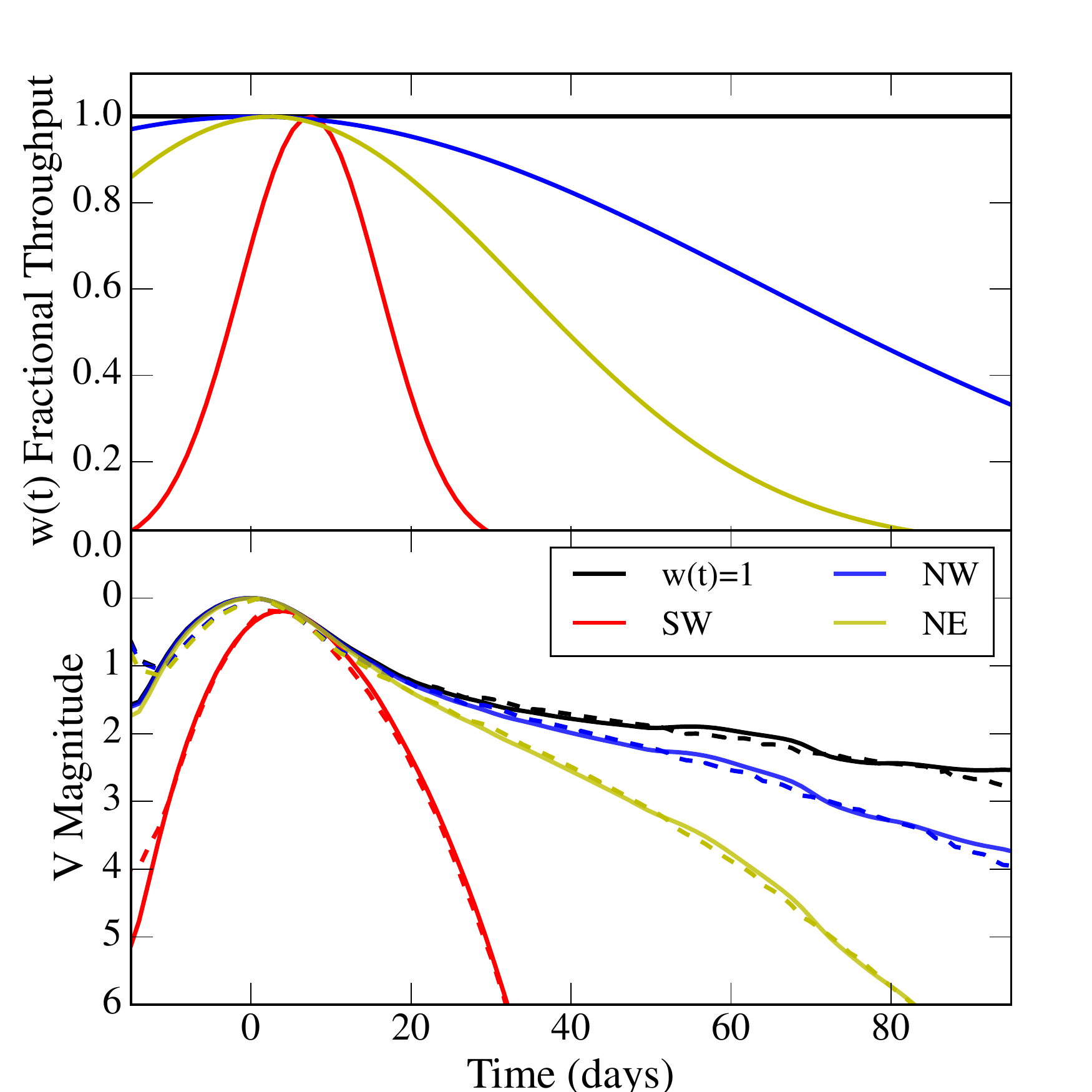}
\caption{Window functions of the three LEs of Cas~A (top panel). Window functions convolved with the SN~IIb template V-band light curve (F. B. Bianco et al., 2016, in prep.), showing how much each epoch contributes to the measured spectrum (bottom panel). The black curve is the unmodified template light curve, which corresponds to a window function $w(t)=1$. This figure should be compared to figure 3 in R11 (our lower panel shows their corresponding curves as dashed lines) and differs only in the use of a template light curve as opposed to that of an individual SN (SN~1993J). The differences are small, but noticeable at early times when SN~1993J shows an additional early peak (due to cooling envelope emission), which is not present in our templates after $t_{Vmax}=-$15 days. \label{fig:window_functions}}
\end{figure}

\citet{Rest:2010yi} calculated the window functions of the three LEs of Cas~A under study using properties of the dust sheet and the observing slit, the distance from the source and the photometric evolution of the transient. The window functions we use in this work are identical in all respects except where photometry is incorporated and are shown in Figure~\ref{fig:window_functions}.
To apply the window functions correctly we need flux as a function of time and wavelength, and spectra from each epoch must be weighted appropriately so that they have the correct relative normalization. However, since the spectra we use only have good relative calibration, but not absolute calibration, we must use the corresponding light curves to infer the correct normalization of each spectrum.
While R11 used light curves of individual SNe (SN~1993J and SN~2003bg) we must use a light curve relevant to the template spectra of L16 that we use. We must therefore use template SNe~IIb light curves derived from the photometric observation of a large sample of SNe~IIb. Such light curves have been created by F. B. Bianco et al. (2016, in prep.). Later, when we use individual SN for the comparison, the spectra and light curves of the specific SN will be used.

SN~1993J, which was used as a reference by R11, has a prominent double peak in its early time photometric data. 
A double-peaked rise structure is present in several SNe~IIb, yet is not universal, and its details are diverse. The double peak is not present in our template for epochs $t_{Vmax}>-15$ days. There is therefore a noticeable difference in the window functions shown in Figure~\ref{fig:window_functions} and those in R11 (shown as dashed lines), despite the minimal difference in construction.

Furthermore, since it is hard to quantify the effect of reddening due to dust in the SN's host galaxy, we cannot trust the shape of the spectra on large wavelength scales. The templates have therefore been constructed using continuum-subtracted spectra (L16) and assumptions must be made in order to account for the missing continuum. Since we will be focusing only on individual features in the spectra, the assumption that we choose to make is that the continuum varies only on very large wavelength scales and hence is flat throughout the range of wavelengths covered by each feature, which is on the order of $200$ \AA. With this assumption reinserting the continuum is trivial and each epoch is simply weighted by a single band light curve, observed in the appropriate photometric band for a given feature. Figure~\ref{fig:features} shows the template spectra as they would appear in the three LEs of Cas~A with each panel focusing on a single feature in the spectrum. For the Ca II NIR triplet and H$\alpha$ the choice of which band to use for normalization is obvious, however the He I $\lambda$5876 line lies directly in between the V and r bands. We therefore use both bands for this line and compare the results. As is evident from this plot, the choice of band is not significant and we therefore use the V band for the remaining analysis.

For some SNe (SN~1993J, SN~2008ax, SN~2011dh and SN~2011ei) the light curves in the literature do not cover the full range of phases for which spectra have been obtained. For these SNe we extrapolate the light curve to late phases using a linear extrapolation in log space (magnitudes) by fitting a linear decline to the SN brightness for $t_{Vmax}>$  50 days, so that we can appropriately normalize all the spectra. A linear extrapolation is consistent with our lightcurve templates as well as with the expectation from $^{56}$Co decline. We are therefore able to include all observed spectra in our calculations. The window functions only have power for ${t_{Vmax}\lesssim100}$ days, as is evident from Figure~\ref{fig:window_functions}, and therefore spectra taken at later phases than this contribute negligibly to the final spectrum. SN~2011dh and SN~1993J have light curves measured to 80 and 94 days after V-max respectively so this extrapolation is of little consequence for these SNe. However, the lightcurves of SN~2011ei and SN~2008ax are only measured to 33 and 50 days respectively and so the extrapolation is more important. For these SNe we use only the last 2 points for the extrapolation, since the slope at earlier phases may not be indicative of the SN's decline at late phases. We notice that the last V-band measurement in the lightcurve of SN~2011ei \citep[figure 3]{Milisavljevic:2012qr} indicates a very shallow decay rate, much slower than would be extrapolated from the earlier datapoints (the last of which is at $t_{Vmax}=19$ days). While we have no reason to doubt this measurement, and some SN~IIb lightcurves do show flattening after 20 days \citep{Prentice:2016ijw}, we felt it was important to test the effect that this datapoint has on our results. We therefore generated two sets of LE spectra for SN~2011ei using two extrapolations of its lightcurve, one including the measurement at 33 days (a very shallow decline) and one ignoring it (a rather steep decline). The only result affected by this choice is the velocity of the He I $\lambda$5876 line seen through the NW LE (whose window function has more power at late times), which changes from -9100$kms^{-1}$ to -9600$kms^{-1}$, a small change that leaves the conclusions of this paper intact.

\subsection{Calculating the Absorption Velocity}
\label{sec:velocity}

The metric we choose for our comparison is the absorption velocities of three features in the spectrum: He I $\lambda$5876, H$\alpha$ and the Ca II NIR triplet, as was done by R11.
We employ a Monte Carlo (MC) method, adapted from the one used in L16, to calculate these velocities, the details of which follow.
We work separately on each feature by choosing a $\sim400$\AA~region around the expected location of the feature and generate synthetic spectra within that region from the observed flux values and their uncertainties. 
We then fit a quadratic 
polynomial to an 80\AA~range around the flux minimum for each realization, and take the location of the minimum of the quadratic fit as our velocity measurement, using the relativistic Doppler formula to convert between wavelength and velocity. For the Cas~A spectra we smooth each synthetic spectrum using the Gaussian process described below before fitting the quadratic since the low SNR of these spectra make direct fits unreliable.


The distribution from which our synthetic observations are drawn should be the same as the one our actual observation was drawn from, i.e. the distribution of observations we would expect were we to return to our telescope on several different nights.
For the template and the directly observed spectra it is a good enough approximation to assume that the observed values are close enough to the true spectral flux to draw synthetic observations from a distribution centered on the observed flux. However, due to their low SNR at native spectral resolution, we cannot generally assume that this holds for the LE spectra. Instead of binning the spectra to suppress the noise, we can calculate the correct distribution using Gaussian processes in the following way.
We first place a prior on the expected spectrum driven by our expectation that SN spectra vary smoothly. We then condition this prior on the measured spectrum to obtain the final distribution from which our synthetic observations, $f_i$ at wavelength $\lambda_i$, are drawn. We choose our prior to be an exponential with mean zero and a covariance of 
\begin{equation}
k(\lambda_1,\lambda_2)=N\exp\left(-\frac{\left(v(\lambda_1)-v(\lambda_2)\right)^2}{\sigma^2}\right)
\end{equation}
between any two fluxes measured at wavelengths $\lambda_1$ and $\lambda_2$, where $v(\lambda)$ is the velocity of wavelength $\lambda$ as measured by the relativistic Doppler formula. $N$ and $\sigma$ are parameters of our model, which represent the SNR and the minimum expected line width respectively. We choose $N=1$ and $\sigma=3\e{3}$km/s, in order to be consistent with the smoothing algorithm used by R11. The mean vector, $\bar{f}$, and covariance matrix, $\cov{f}$, of the final distribution from which our synthetic observations are drawn are then
\begin{eqnarray}
\bar{f}&=&K(K+\Epsilon^2)^{-1}f_*\\
\cov{f}&=&K+\Epsilon^2-K(K+\Epsilon^2)^{-1}K,
\end{eqnarray}
where $K$ is the covariance matrix of our prior, with elements ${K_{ij}=k(\lambda_i,\lambda_j)}$; $\Epsilon$, with elements ${\Epsilon_{ij}=\epsilon_i\delta_{ij}}$, is a diagonal matrix of the observational errors of our measurements and $f_*$ is the observed flux. Full details of this calculation can be found in e.g. \citet{rasmussen2006gaussian}.

We then generate $N_{MC}=N_\lambda(\ln N_\lambda)^2$ simulations where $N_\lambda$ is the number of wavelength bins within the range of interest \citep{babu1983}\footnote{This is proven to be an appropriate number of draws for bootstrapping. By extension we assume it is an appropriate number of draws for our MC simulations.}.
We smooth each simulated spectrum by the matrix ${K(K+\Epsilon^2)^{-1}}$, treat it as an independent observation and process it as we did for the directly observed spectra.

\begin{figure*}
\includegraphics[width=\textwidth]{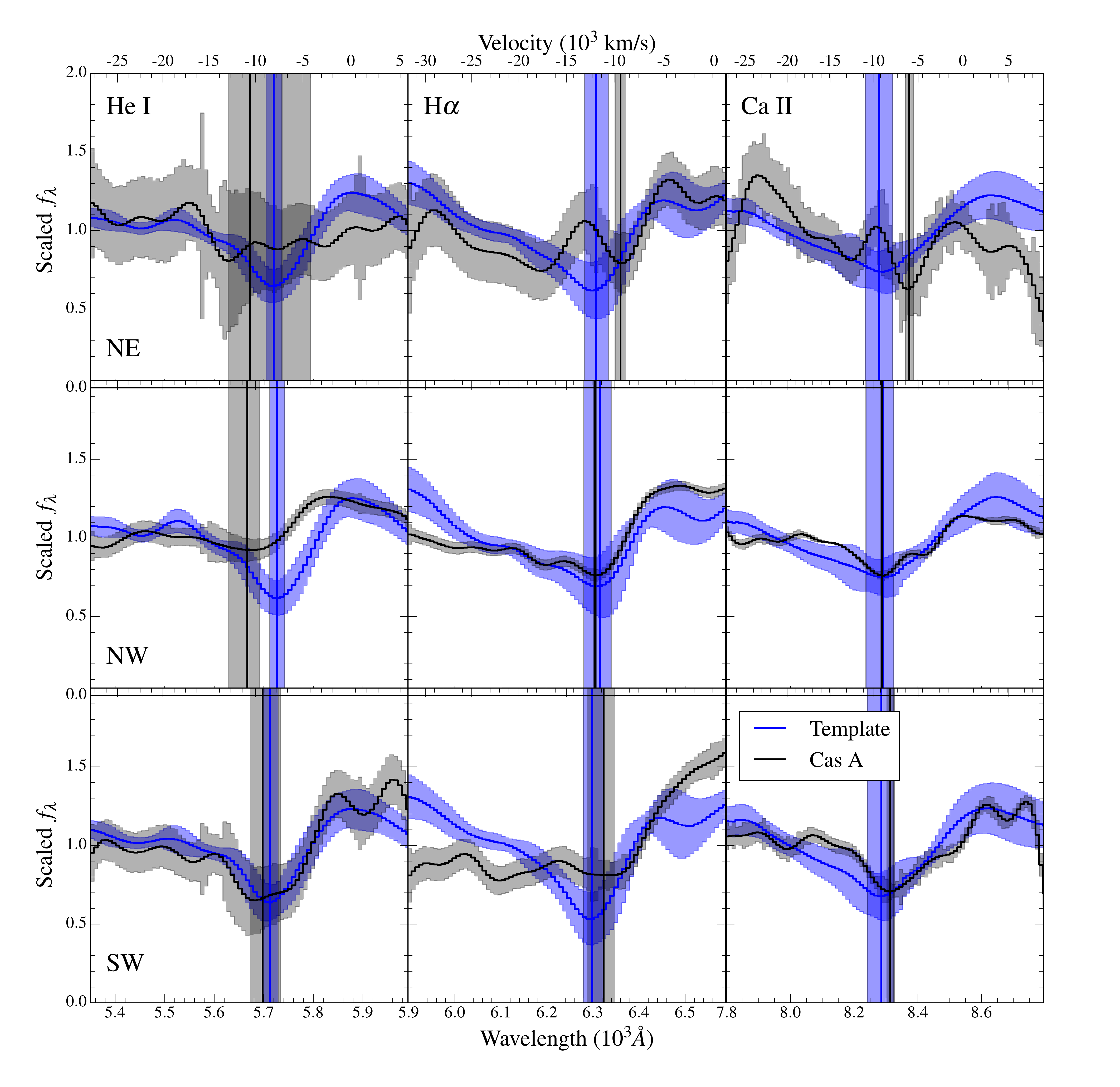}
\caption{Comparison of the mean template spectra, processed through the window functions of the three dust sheets (blue solid lines), with the observed spectra of the LEs of Cas~A (black solid lines). The shaded light blue bands represent the 1$\sigma$ variance in the template spectra and the grey bands represent the 1$\sigma$ observational uncertainty in the Cas~A spectra. The observational uncertainty increases dramatically at the location of telluric and night sky lines (for example the OI night sky line at 5577\AA). Each column displays the same spectral feature: He I $\lambda$5876, H$\alpha$, and the Ca~II NIR triplet from left to right, and each row displays a light echo: NE, NW and SW from top to bottom. The positions of the minima of the relevant feature, as calculated using the MC techniques, described in the text are shown as vertical lines for the median values with vertical bands representing the 1$\sigma$ uncertainty.  We see that, for all lines, the velocity in the NW direction is larger than that of the other two directions, confirming R11's result that Cas~A is asymmetric. 
  \label{fig:casa_mins}}
\end{figure*}

Figure~\ref{fig:casa_mins} shows the range of velocities found superimposed on the Cas~A LE spectra and the template spectra processed through the appropriate window functions.

\section{Analysis of Asymmetry and Diversity}
\label{sec:results}
\subsection{Confirming the Asymmetry of Cas~A}
By comparing the velocity of our three absorption features in the LE spectra of Cas~A to those of the template spectra convolved with the relevant window functions of the dust sheets we can isolate effects due to the asymmetry of Cas~A from effects due to differences in the dust sheets. If Cas~A were symmetric we would expect the velocity of Cas~A to be the same in all three directions with respect to the template, i.e. we would expect either the line velocities from all three directions to be consistent with the templates or for them all to be faster or all slower by the same amount. Any deviation from this is indication of asymmetry.

As can be seen in Figure~\ref{fig:casa_mins} for all three lines the highest velocity, relative to the template velocity, is observed in the NW direction.
For the H$\alpha$ line we see that Cas~A is consistent with the template when seen from the NW direction but has a lower velocity than the templates, with significance $2\sigma$, when seen from the NE direction and $1\sigma$ when seen from the SW direction. Similarly for the Ca II NIR triplet Cas~A is perfectly consistent with the template when viewed from the NW direction, whereas it is just on the slow end of 1$\sigma$ consistency in the SW direction and is slower by $2\sigma$ in the NE. The He I $\lambda$5876 line has the largest uncertainty of the three so care must be taken when considering this line. Nevertheless we see that Cas~A has higher velocity than the template, with a significance of $\sim2\sigma$ when viewed from the NW direction whereas it is consistent at $1\sigma$ when viewed from the SW 
and NE directions. 
The uncertainty in the NE LE in particular is very large for this line, thus although the velocity is consistent with the templates at $1\sigma$, this has little impact on the significance of our conclusions.

Although for no individual line we find a statistically significant difference in velocity by itself, the fact that all three lines show NW as the fastest moving direction, as well as the other evidence for asymmetry of Cas~A mentioned in Section~\ref{sec:intro} leads us to the same conclusion as R11: that Cas~A is indeed asymmetric and, in particular, the ejecta is moving more quickly in the NW direction. Further, our calculations strengthen this result since we use mean SN~IIb spectra instead of individual SNe and employ MC methods to quantify the differences in velocity more robustly.

\begin{figure*}\center{
		\subfigure{\includegraphics[width=\textwidth]{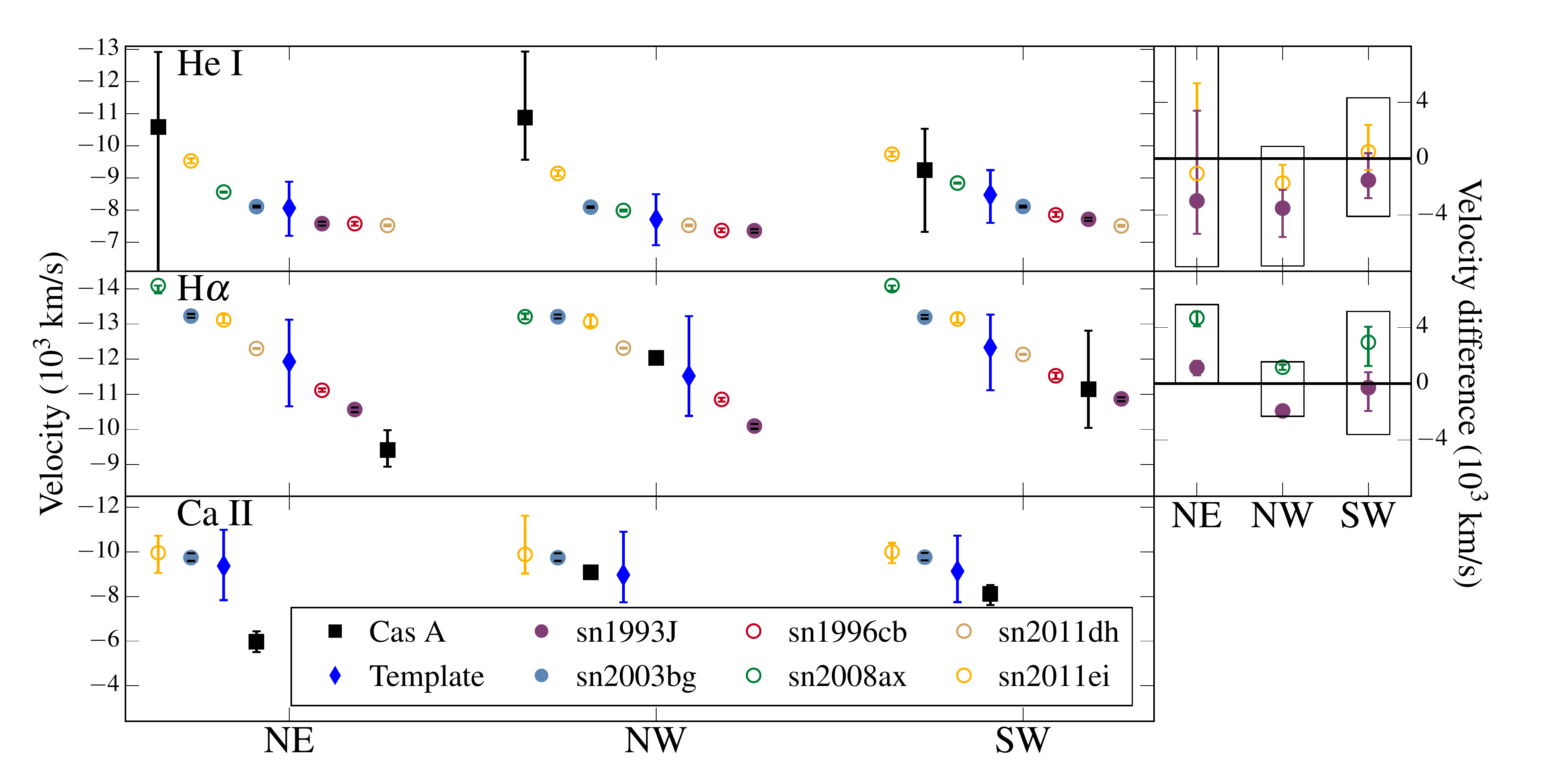}}}
        \caption{Comparison of the absorption velocity of directly observed SNe~IIb, processed through the window functions of the relevant dust sheets, with the measured velocity of Cas~A in those LEs. The error bars for the reference SNe are included, but in many cases lie within the markers. In the right panel the velocity difference between the fastest SN and Cas~A, and between the slowest SN and Cas~A, are plotted for each light echo for the He~I $\lambda$5876 (top) and H$\alpha$ lines (bottom). The box around each set of measurements represents the $2\sigma$ confidence region for the velocity difference, including the errors on Cas~A and on the SN measurements.  
The Cas~A velocities span the complete range of diversity seen in SNe~IIb with some laying at the high end of the distribution (e.g. NW He~I $\lambda$5876) and others at the low end (e.g. NE H$\alpha$) supporting our hypothesis that at this stage the observed diversity can be explained purely through asphericity. Only two of the SNe~IIb in our sample had enough NIR spectral data to enable a reliable simulation of the effects of LE reflection. Therefore the objects shown for this line do not represent the full diversity of SNe~IIb, and the line is omitted from the right panel. \label{fig:velocity}}
\end{figure*}

\begin{figure}\center{
        \includegraphics[width=\columnwidth]{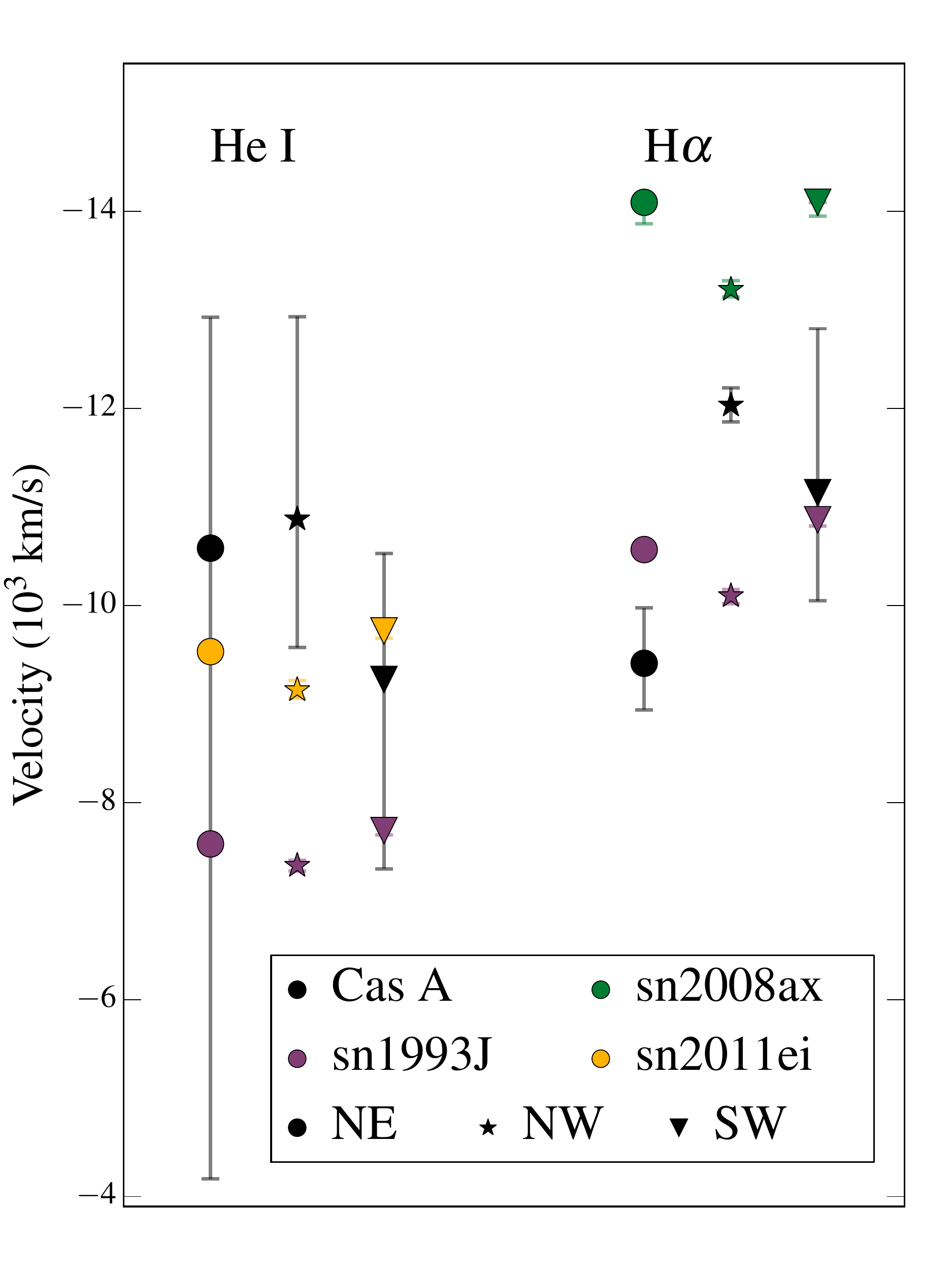}
        \caption{Comparison of the absorption velocity of the fastest, and slowest SN from our SNe~IIb sample processed through the window functions of the relevant dust sheets with the measured velocity of Cas~A in those LEs. These are the same data shown in Figure~\ref{fig:velocity}, but grouped by SN and only showing those SNe with the most extreme velocities. The error bars for the reference SNe are included, but in many cases lie within the markers. The effects of the reflection in the dust, simulated through the application of the appropriate window function, determines the difference in the three velocities of each SN. Notice that the effect of dust reflection from the NW light echo decreases the velocity in the directly observed SN, but for Cas~A spectra from this light echo shows the fastest velocity, proving that material is ejected at higher speed in this direction.}\label{fig:velocity2}}
\end{figure}

\subsection{Comparison with the SN~IIb Sample}
In order to see if this asymmetry, as seen in the Cas~A LE spectra, is enough to explain the diversity of absorption velocities seen in our sample of SNe~IIb we repeat the same steps as before, but use the spectra of the individual SNe in our sample as references instead of the template. Figure~\ref{fig:velocity} shows the absorption velocities, as measured by our MC technique, of our selection of SNe~IIb processed through the window functions of the three dust sheets, as well as the LEs of Cas~A. These measurements are also tabulated in Table~\ref{tab:velocities}. The sample of directly observed SNe~IIb includes SNe with the most extreme velocities as identified by L16 
(see Figure~\ref{fig:yuqian_velocity}), and for the H$\alpha$ and He~I $\lambda$5876 lines, measurements of the fastest and slowest well observed SN in the sample are also plotted separately in Figure~\ref{fig:velocity2}.



For both the He I $\lambda$5876 and H$\alpha$ lines, the velocity of Cas~A is consistent to within $2\sigma$ with the velocity of either the fastest, or slowest SN in our sample in at least one LE. Because of this we cannot rule out that the diversity observed in SNe~IIb, as measured from the blue-shift velocity of the lines, is entirely due to asymmetry in the explosion, while the kinetic energy per unit mass could be identical for all objects in the sample. Meanwhile, diversity in the kinetic energy per unit mass may of course manifest itself in other observables, for example in luminosity, line ratios, or temperatures, which our analysis does not probe.

Again we notice that the velocity difference between the different LEs of Cas~A is greater than that of any other SN in our sample, especially in H$\alpha$, again highlighting that the asymmetry of Cas~A is real, and not just an artifact of the differences in the dust sheets (while in our sample of SNe~IIb each object is, of course, seen only from a single line-of-sight, and all differences in the different panels arise from the convolution with different LE window functions).

In Figure~\ref{fig:velocity2} only the SNe with the most extreme velocities are plotted, and the velocities are grouped by SN. Data points with the same symbol, representing the same LE, should be compared, and we see the following.
\begin{itemize}
\item{While the velocities of the reference SNe are only slightly affected by the different window functions, a big difference between the LEs of Cas~A is seen, which we attribute to to a direct detection of asymmetry in Cas~A.  This effect is even more apparent when one considers that, for most of the reference SNe, the window function of the NW dust sheet causes the feature to appear slower, whereas for Cas~A, the velocity in this direction is consistently the highest of the three (Figure~\ref{fig:velocity2}).}
  
\item{
Some of the velocities of Cas~A are consistent with the fastest SNe IIb (e.g. NW He I $\lambda$5876) whereas others are consistent with the slowest (e.g. NE H$\alpha$).
Thus the range of velocities seen in the LEs of Cas~A is of the same order as the spread of velocities in our sample of SNe~IIb, indicating that the asymmetry of Cas~A, if common among SNe~IIb, would manifest itself as diversity in a sample of directly observed SNe~IIb consistent with that which we observe.}

\item{For all lines and all directions there is at least one SN~IIb whose velocity difference from Cas~A is consistent with $0$ within $2\sigma$ (see the right panel of Figure~\ref{fig:velocity}). This shows that Cas~A is not an outlier of the distribution of velocities in SNe IIb and hence that treating the asymmetry of Cas~A as typical is a justified assumption.}

\item{The SNe~IIb displaying the most extreme velocities in He I $\lambda$5876 are not the same objects that show extreme velocities in H$\alpha$. In fact for all LEs, Cas~A has high He I $\lambda$5876 velocities, but low $H\alpha$ compared to our sample and our template. }
\end{itemize}

\begin{deluxetable*}{cccccccccc}
\tabletypesize{\footnotesize}
\tablecolumns{10}
\tablewidth{0pt}
\tablecaption{Line velocities, in units of $10^3kms^{-1}$, as seen through the LEs, for different SNe~IIb. $1\sigma$ uncertainties are shown in parentheses.\label{tab:velocities}}
\tablehead{&&\colhead{Cas A}&\colhead{Template}&\colhead{sn2003bg}&\colhead{sn2011ei}&\colhead{sn1993J}&\colhead{sn1996cb}&\colhead{sn2008ax}&\colhead{sn2011dh}
}
\startdata
He I&&&&&&&&&\\
&NE&-10.58 (-4.37)&-8.06 (-0.84)&-8.11 (-0.03)&-9.53 (-0.08)&-7.58 (-0.06)&-7.58 (-0.06)&-8.56 (-0.02)&-7.52 (-0.03)\\
&NW&-10.88 (-1.68)&-7.71 (-0.79)&-8.09 (-0.03)&-9.14 (-0.09)&-7.36 (-0.05)&-7.37 (-0.06)&-7.99 (-0.03)&-7.53 (-0.03)\\
&SW&-9.25 (-1.60)&-8.48 (-0.82)&-8.11 (-0.03)&-9.74 (-0.08)&-7.72 (-0.05)&-7.85 (-0.08)&-8.84 (-0.02)&-7.51 (-0.03)\\
H$\alpha$&&&&&&&&&\\
&NE&-9.41 (-0.52)&-11.93 (-1.23)&-13.23 (-0.05)&-13.12 (-0.12)&-10.57 (-0.07)&-11.11 (-0.05)&-14.09 (-0.11)&-12.30 (-0.02)\\
&NW&-12.03 (-0.17)&-11.52 (-1.42)&-13.21 (-0.06)&-13.06 (-0.17)&-10.10 (-0.07)&-10.86 (-0.06)&-13.21 (-0.08)&-12.31 (-0.02)\\
&SW&-11.15 (-1.38)&-12.33 (-1.08)&-13.19 (-0.05)&-13.14 (-0.14)&-10.87 (-0.05)&-11.53 (-0.09)&-14.09 (-0.07)&-12.13 (-0.02)\\
Ca II&&&&&&&&&\\
&NE&-5.98 (-0.47)&-9.37 (-1.58)&-9.74 (-0.17)&-9.96 (-0.84)&&&&\\
&NW&-9.09 (-0.12)&-8.97 (-1.58)&-9.74 (-0.17)&-9.89 (-1.30)&&&&\\
&SW&-8.12 (-0.44)&-9.14 (-1.49)&-9.77 (-0.16)&-10.01 (-0.45)&&&&\\
\enddata

\end{deluxetable*}

\section{Conclusions}
\label{sec:conc}
In this work we have re-analyzed the light echo spectra of Cas~A discovered and studied by R11 in order to quantify its asphericity, as traced by spectral line velocities seen from different directions. We also tested the hypothesis that asphericity, as measured in Cas~A, is enough to
explain the diversity of spectral line velocities seen in extragalactic SNe IIb (in each case seen only along one line of sight).

We have confirmed the result of R11 by showing that the velocities of the He I $\lambda$5876, H$\alpha$ and Ca II NIR triplet lines in Cas~A are all larger when viewed from the NW direction, even after considering the effect of the dust sheets that produce the LEs by reproducing the effects in reference spectra. We have strengthened R11's result by using mean SN~IIb spectra as a reference, as well as several individual SNe~IIb, thereby showing that this result is not sensitive to the choice of reference. We have determined uncertainties on our velocity measurements of template, individual SNe, and light echo spectra through MC methods so that we can make the comparisons more robustly and can assess the statistical significance of the result we present.

The range of velocities due to the asymmetry of Cas~A is of the same order as the range observed in the population of SNe~IIb and thus the velocity diversity in the observed sample of SNe~IIb is consistent with being induced by asymmetry, even in the presence of explosions with identical kinetic energy per unit mass that are asymmetric to the level of Cas~A. We thus cannot rule out the possibility that the velocity diversity seen in SNe~IIb is due to their intrinsic asymmetry, especially when we consider the possibility that there is another direction moving even faster than NW, or slower than SW and NE, but which doesn't have a corresponding observed LE. Our result indicates that asymmetry has a significant contribution in determining the observed diversity of spectral velocities within SNe~IIb. Observations of other SNe~IIb with multiple LEs would be the best way to strengthen this result, since currently there is no way to exclude the possibility that Cas~A is more asymmetric than a typical SN~IIb. Additional objects could confirm whether asymmetry in SNe~IIb is common, and quantify the contribution of asymmetry to the spread in line velocities seen in the directly observed SNe~IIb. The analysis in this paper has been limited to the diversity in line velocities, however other measurements, such as line flux ratios or temperature, may also indicate additional diversity, which we do not probe here.

Another interesting observation
is that the SNe that are most extreme in one spectral feature are not necessarily as extreme in another. In particular, SN~2011dh has one of the slowest He I $\lambda$5876 lines in our sample, yet in the H$\alpha$, it is average, or even among the faster SNe. This may indicate that different envelopes of the SN move at different velocities. Cas~A itself shows high He I $\lambda$5876 velocities, but lower $H\alpha$ velocities,  for all LEs, when compared to the directly observed SNe.

Our result is limited by the small sample size for both LEs and well observed SNe~IIb. This of course will change in the near future: modern and future synoptic surveys, starting with PTF \citep{Law09,Rau09}, PanSTARRS \citep{Tonry12}, DECAM \citep{DECAM}, ZTF \citep{Smith12}, and culminating with LSST \citep{LSST}, are discovering SNe at an ever increasing rate. Studies like ours require dense spectroscopic time series, which will be limited to the closest few objects, but it is undoubtable that our understanding of the diversity of SNe~IIb and other SN subtypes will dramatically improve in the years to come. LEs as well will be discovered at a much higher rate in the LSST age, with the increased sky coverage at faint magnitude. Here too we will need the largest telescopes to collect spectra of the low surface brightness LEs, but the sample of SNe with direct asymmetry constraints should increase. This will not only allow us to reproduce this study and obtain a more statistically significant result, but also to extend this investigation of how asymmetry contributes to observed diversity, and how intrinsically diverse SNe that are classified within the same subtype are, to other SN classes. For example, while LEs of the SNR Tycho have been observed \citep{Rest:2007ab}, which indicate that it was a SN~Ia \citep[e.g.][]{Rest:2008eg,Krause:2008bn}, it would be crucial to obtain spectra of LEs that arose from different viewing angles in order to constrain the degree of asphericity in SNe~Ia. This could help constrain their progenitors since, for example, the head-on collision of white dwarfs in a triple system \citep{Kushnir:2013ab} claims large asphericities \citep{Dong:2014ab}. 
\section{Acknowledgments}
We are grateful to Robert Speare, Michael Blanton, Saurabh Jha and Or Graur for useful discussions and Niloufar Khavari for preliminary work on this project. We thank the referee for helpful suggestions that improved the quality of the paper.

K. Finn is supported in part by the \emph{US-UK Fulbright Commission All-Disciplines Postgraduate Award}.
F. B. Bianco is supported in part by the \emph{NYU/CCPP James Arthur Postdoctoral Fellowship}.
M. Modjaz is supported in parts by the NSF CAREER award AST-1352405 and by the NSF award AST-1413260.
Y. Q. Liu is supported in part by the NSF award AST-1413260 and by a \emph{NYU/CCPP James Arthur Graduate Award}.
A. Rest was in part supported by the HST programs GO-12577 and AR-12851.


\end{document}